\documentclass{aa}  

\usepackage{graphicx}
\usepackage{float}

\usepackage{txfonts}
\usepackage{hyperref}  
\hypersetup{colorlinks=true,linkcolor=[rgb]{1.,0.2,0.2},citecolor=[rgb]{0.1,0.4,1.},filecolor=[rgb]{0.7,0.2,0.2},urlcolor=[rgb]{0.7,0.2,0.2}}

\usepackage{color}
\definecolor{blue}{rgb}{0., 0., 1}
\newcommand {\T}{Table\,}
\newcommand {\Sec}{Sec.\,}
\newcommand {\Fig}{Fig.\,}
\newcommand {\Eq}{Eq.\,}

\begin{document}

   \title{Probing diffuse radio emission in bridges between galaxy clusters with uGMRT
}

   \author{G.V. Pignataro
          \inst{1,2} \fnmsep\thanks{e-mail: \href{mailto:giada.pignataro2@unibo.it}{giada.pignataro2@unibo.it}}
          \and
          A. Bonafede\inst{1,2}
          \and 
          G. Bernardi\inst{2,3,4}
          \and
          C.~J. Riseley\inst{1,2,5}
          \and
          D. Dallacasa\inst{1,2}
          \and
          T. Venturi\inst{2,3}
          }

   \institute{Dipartimento di Fisica e Astronomia, Universit\`a degli Studi di Bologna, via P. Gobetti 93/2, 40129 Bologna, Italy
        \and
             INAF -- Istituto di Radioastronomia, via P. Gobetti 101, 40129 Bologna, Italy
        \and
            Department of Physics and Electronics, Rhodes University, PO Box 94, Makhanda, 6140, South Africa
        \and
            South African Radio Astronomy Observatory (SARAO), Black River Park, 2 Fir Street, Observatory, Cape Town, 7925, South Africa
        \and 
            CSIRO Space \& Astronomy, PO Box 1130, Bentley, WA 6102, Australia
            }

   \date{Received September 15, 1996; accepted March 16, 1997}

 
  \abstract
   {}
   {Recent X-ray and Sunyaev-Zeldovich (SZ) observations have detected thermal emission between early stage merging galaxy clusters. The main purpose of this work is to investigate the properties of the non-thermal emission in the interacting clusters pairs Abell 0399-Abell 0401 and Abell 21-PSZ2 G114.9. 
}
   {These two unique cluster pairs have been found in an interacting state. In both cases their connection along a filament is supported by SZ effect detected by the \textit{Planck} satellite and, in the special case of Abell 0399-Abell 0401, the presence of a radio bridge has been already confirmed by LOFAR observations at 140~MHz. Here, we analyse new high sensitivity wideband (250-500~MHz) uGMRT data of these two systems and describe an \textit{injection} procedure to place limits on the spectrum of Abell 0399-Abell 0401 and on the radio emission between Abell 21-PSZ2 G114.9. 
}
   {In both cases, the low-surface brightness diffuse emission is not detected in Band3 (250-500~MHz). For the A399-A401 pair, we are able to constrain the steep spectral index of the bridge emission to be $\alpha>2.2$ with a 95\% confidence level between 140~MHz and 400~MHz. We also detect a small patch of the bridge with a flatter spectral index, that may suggest a variable spectral index distribution across the bridge area. For the A21-PSZ2 G114.9 pair, we are able to place an upper limit on the flux density of the bridge emission with two different methods, finding at the central frequency of 383~MHz a conservative value of $f_{u}^{1}<260$~mJy at 95\% confidence level, and a lower value of $f_{u}^{2}<125$~mJy at 80\% confidence level, based on visual inspection and a morphological criterion.
}
   {Our work provides a constraint on the spectrum in the bridge A399-A401 which disfavours shock-acceleration as the main mechanism for the radio emission. The methods that we propose for the limits on the radio emission in the A21-PSZ2 G114.9 system represent a first step toward a systematic study of these sources.}

   \keywords{
               }

   \maketitle
%

\section{Introduction}
The large-scale structure of the Universe can be described as web-like pattern, where low-density filaments of gas connect knots associated to galaxy clusters. The accretion of matter onto galaxy clusters happens along the filaments of the so called cosmic-web \citep{bond1996}, and the subsequent merger of these systems releases an extreme amount of energy into the intracluster medium (ICM). If part of this energy is channeled into particle acceleration and magnetic field amplification we expect to observe synchrotron radio emission on Mpc-scales. 

The evidence of such processes in galaxy clusters is represented by different types of radio diffuse sources in the ICM: radio relics, mini halos and radio halos. Radio relics and radio halos both extend for $\sim$Mpc-scales and they are found in dynamically disturbed clusters, located in the peripheries and central regions, respectively. They are the signposts of the (re-)acceleration of electrons through shock processes, for relics \citep[e.g.,][]{hoeft2007}, and turbulence, for halos \citep[e.g.,][]{brunetti2016}, generated by the merger event. Mini-halos on the other hand, are found in the core of relaxed clusters and they could be generated by turbulence created by gas-sloshing \citep{zuhone2013}, or by hadronic collisions \citep{Pfrommer04}. They are all characterised by steep radio spectrum\footnote{$S_{\nu}\propto \nu^{-\alpha}$} ($\alpha>1$) and low-surface brightness at high frequencies \citep[for a recent review, see][]{vanweeren2019}.

\begin{figure*}[h!]
    \centering
    \includegraphics[width=1\linewidth]{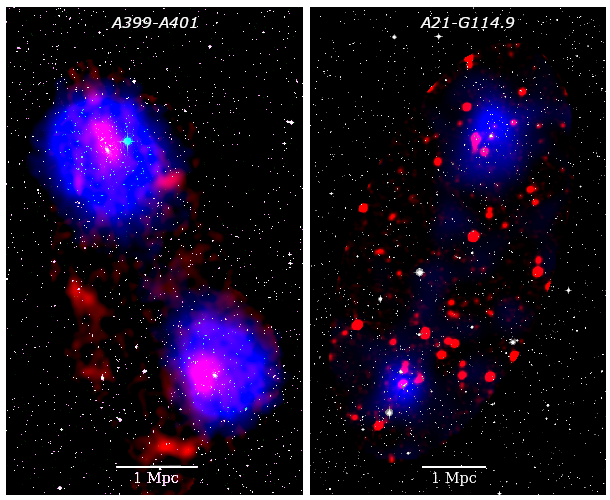}
    \caption{Composite image of the A399-A401 (\textit{left}) and A21-G114.9 (\textit{right}) cluster pairs. Optical data are recovered from the DSS, while X-ray data (ASCA for A399-A401, ROSAT for A21-G114.9) are shown in blue, and uGMRT data from this work are overlaid in red.}
    \label{fig:rgb}
\end{figure*} 

Recent low-frequency observations have shown the presence of diffuse radio emission on even larger scales, along the filaments of the cosmic-web between interacting cluster pairs \citep{govoni2019, botteon2020, hoeft2021}. In addition, radio bridges are being discovered also between clusters and groups of galaxies, as in the cases of the Coma cluster, detected at low-frequency \citep{bonafede2021}, and, for the first time at high-frequency, in the Shapley Supercluster \citep{venturi2022}. The presence of non-thermal emission on the cosmic large-scale (1-15~Mpc) is reported also in \cite{vernstrom2021}, where they find a robust detection of stacked radio signal from the filaments between luminous red galaxies. Multi-frequency studies of synchrotron emission from radio-bridges between clusters are fundamental to shed light on mechanisms of particle acceleration and properties of the magnetic fields on scales never probed before \citep{vazza2019}. The discovery of such bridges stressed the need to find theoretical model that can explain this emission, different than radio relics and halos. \cite{govoni2019} explored different possibilities to explain their observations, since the synchrotron and inverse Compton losses make the lifetime of the particles ($\sim 10^{8}$~years at 140~MHz) too short to travel from the center of the cluster and cover the bridge extension. This points to an \textit{in-situ} mechanism for particle acceleration, such as diffuse shock re-acceleration of a pre-existing population of mildly relativistic electrons. This process would plausibly result in a spectral index of $\alpha \sim 1.2 - 1.3$ for the bridge, as often observed in relics \citep{vanweeren2019}. Recently, \cite{brunetti2020} presented a model that could explain the origin of radio bridges as synchrotron emission from fossil seed particles (from past AGN or star-formation activity) re-accelerated in turbulence generated along the filament of accreting compressed matter. The resulting emission should be characterised by steep spectrum ($\alpha>1.3$). Therefore, to test the models it is important to characterise the spectral properties of these structures. 

\begin{table}[h!]
\centering
\resizebox{\columnwidth}{!}{%
\begin{tabular}{lllll}
\hline \hline
\multicolumn{1}{c}{Target} & \multicolumn{1}{c}{\begin{tabular}[c]{@{}c@{}}Right Ascension \\ (J2000)\end{tabular}} & \multicolumn{1}{c}{\begin{tabular}[c]{@{}c@{}}Declination \\ (J2000)\end{tabular}} & \multicolumn{1}{c}{Redshift} & \multicolumn{1}{c}{\begin{tabular}[c]{@{}c@{}}Mass \\ $(M_{\odot})$\end{tabular}} \\ \hline
A399                       & 02h 57m 56s                                                                           & +$13^{\circ}$00$^{\prime}$ 59$^{\prime\prime}$                                                            & 0.072                        & $\sim 5.7 \times 10^{14}$                                                         \\
A401                       & 02h 58m 57s                                                                           & +$13^{\circ}$34$^{\prime}$ 46$^{\prime\prime}$                                                             & 0.074                        & $\sim 9.3 \times 10^{14}$                                                         \\ \hline
A21                        & 00h 20m 52s                                                                           & +$28^{\circ}$30$^{\prime}$ 30$^{\prime\prime}$                                                             & 0.094                        & $\sim 3.8 \times 10^{14}$                                                         \\
G114.9                    & 00h 21m 13s                                                                           & +$28^{\circ}$15$^{\prime}$ 00$^{\prime\prime}$                                                             & 0.095                        & $\sim 2.5 \times 10^{14}$                                                         \\ \hline
\end{tabular}%
}

\caption{Position, redshift and mass of the two pairs of target galaxy clusters analyzed in this work. For the A399-A401 pair, the mass is X-ray derived by \textit{Chandra} \citep{vikhlinin2009}, while for A21-G114.9 is the SZ derived value \citep[as defined in][]{planck2013}.}
\label{tab:a399a401dets}
\end{table}
The first example of radio bridges is the detection reported in \cite{govoni2019}, between the galaxy clusters Abell 0399 and Abell 0401 (hereafter, A399 and A401). They find a bridge of emission with surface brightness of $I=0.38$~$\mu$Jy~arcsec$^{-2}$ at 140~MHz with LOw Frequency ARray (LOFAR), extended for approximately $3$~Mpc, which is the entire projected separation of the two clusters. An analogous study is presented in \cite{dejong2022}, where they carry out an analysis on the morphology and origin of synchrotron emission in A399-A401 based on a deep 40~hours LOFAR observation.
This local (see \T\ref{tab:a399a401dets}, \citealt{oegerle2001}) system is rich in examples of diffuse emission:  both clusters host a radio halo, detected at high (1.4~GHz, \citealt{murgia2010}) and low-frequency, and some diffuse features possibly classified as radio relics \citep{govoni2019}. The pair is in a pre-merger state \citep{bonjean2018} and X-ray observations \citep{fujita1996, fujita2018, akamatsu2017} revealed the presence of a $6-7$~keV ionised plasma in the region between the clusters. This connection is further supported by the detection of the Sunyaev-Zeldovich (SZ) effect with \textit{Planck} \citep{planck2013,planck2016,bonjean2018, hincks2022} from the gas in the bridge with a density of $\sim 10^{-4}$~cm$^{-3}$.

 Low-frequency radio observations of this cluster pair were carried out with the Westerbork Synthesis Radio Telescope (WSRT) at 346~MHz, but they were not sufficiently deep to detect the radio halo in A401 and the bridge diffuse emission, placing a lower limit on its spectral index at $\alpha>1.5$ \citep{nunhokee2021}.

\cite{bonjean2018} also reported an SZ detection in between another pair of galaxy clusters, Abell 21 and PSZ2 G114.9 (hereafter, A21 and G114.9), separated by a projected distance of approximately 4~Mpc. The morphology of the SZ emission suggests that this nearby pair (see \T\ref{tab:a399a401dets} for details) is found in an interacting, early stage of merger state as well. So far, these two galaxy clusters pairs are unique systems where the \textit{Planck} satellite \citep{planck2013,planck2016} have shown a significant SZ detection in their inter-cluster region. 

In this paper we present high-sensitivity observations with the upgraded Giant Meterwave Radio Telescope (uGMRT) in Band3 (250-500~MHz) of the A399-A401 and A21-G114.9 pairs to investigate the non-thermal properties of their connecting filaments. This work is organized as follows: in \Sec\ref{sec:obsanddata} we describe the data reduction and imaging parameters; in \Sec\ref{sec:a399res} we present the results and discussion on the A399-A401 pair; in \Sec\ref{sec:results_a21psz} we show the results on the A21-G114.9 pair. Throughout this work we assume a $\Lambda$CDM cosmology, with $H_{0}=70$ km s$^{-1}$~Mpc$^{-1}$, $\Omega_{m}=0.3$, and $\Omega_{\Lambda}=0.7$. With these assumptions, at the average distance of the A399-A401 system, $1^{\prime}$ corresponds to $83$~kpc and the luminosity distance is $D_{L}= 329$~Mpc, while at the average distance of the A21-G114.9 system 
$1^{\prime} = 105$~kpc and the luminosity distance is $D_{L}= 360$~Mpc.

\section{Observations and data reduction}\label{sec:obsanddata}

Observations of A399-A401 and A21-G114.9 were carried out with the upgraded GMRT (uGMRT) in band3 (proposal code: $36\_043$, P.I.: Bernardi). The total length of observation is 10~hours per pair, including the time spent on  calibration sources. Each cluster pair is observed with two distinct pointings, one centered on each galaxy cluster. This results in 
approximately 4~hours on-source time for each target galaxy cluster, and approximately 1~hour total spent on calibrators - see \T~\ref{tab:ugmrt_obs} for observational details.

Data reduction was carried out with the \texttt{Source Peeling and Atmospheric Modeling} \citep[\texttt{SPAM};][]{intema2014} pipeline \citep[as described by][]{intema2017_tgss-adr}. The pipeline starts with a pre-processing part, that converts the data into a pre-calibrated visibility dataset by performing several rounds of flagging visibilities affected by radio frequency interference (RFI), and then transferring the calibration solutions derived from the primary calibrator to the data. This is followed by direction-independent calibration on the pre-processed visibilities, with several rounds of phase self-calibration, amplitude self-calibration, and more RFI flagging. Finally, from the resulting self-calibration gain table and final wide-field image, starts the direction-dependent (DD) calibration, which determines the DD gain phases from the peeling of the apparently brightest sources in the field. The gain phases from the peeled sources are spatially fit to constrain a model of the ionosphere, used to predict ionospheric phase delays for arbitrary positions within the field of view.
The total bandwidth is reduced at the start of the data calibration with RFI flagging that include the first and last channels of the band. Then, the remaining wide (200~MHz) band observations are split in six sub-bands, 33.3~MHz each, and the pipeline runs independently on each sub-band. The calibrated data of each sub-band are then jointly imaged with \texttt{WSClean v3.1} \citep{offringa2014}. For the A399-A401 pair, the central frequency of the images is 400~MHz. For the A21-G114.9 pair, we excluded the high frequency (467-500~MHz) sub-band, where \texttt{SPAM} could not find enough sources to fit the ionospheric model. This resulted in an image with rms noise 5 times higher than the other sub-bands. After excluding this sub-band, the remaining calibrated data are imaged at the central frequency of 383~MHz. 

\begin{table}[h!]
\centering
\begin{tabular}{lll}
                 &  &      \\ \hline \hline
Date             &  & 24-25 Aug 2019   \\
Frequency band   &  & $250-500$ MHz \\
N.o. channels    &  & $2048$        \\
Channel width    &  & $97.7$ kHz    \\
Integration time &  & $16.1$ s      \\
Time on source   &  & $4$ hrs       \\
Calibrators      &  & 3C48, 3C147, 3C468.1, \\
                 &  & 2310+110, 0321+123 \\
Correlations     &  & RR, LL        \\ \hline
\end{tabular}
\smallskip
\caption{uGMRT observation details for A399-A401 and A21-G114.9 target clusters pairs. Each observation is comprised of two different pointings, one centered on each galaxy cluster of the pair. The on-source time refers to the single-pointing time spent on each galaxy cluster.}
\label{tab:ugmrt_obs}
\end{table}

\subsection{Imaging and linear mosaicking}\label{sec:imaging}
\begin{figure*}
    \centering
    \includegraphics[width=1\linewidth]{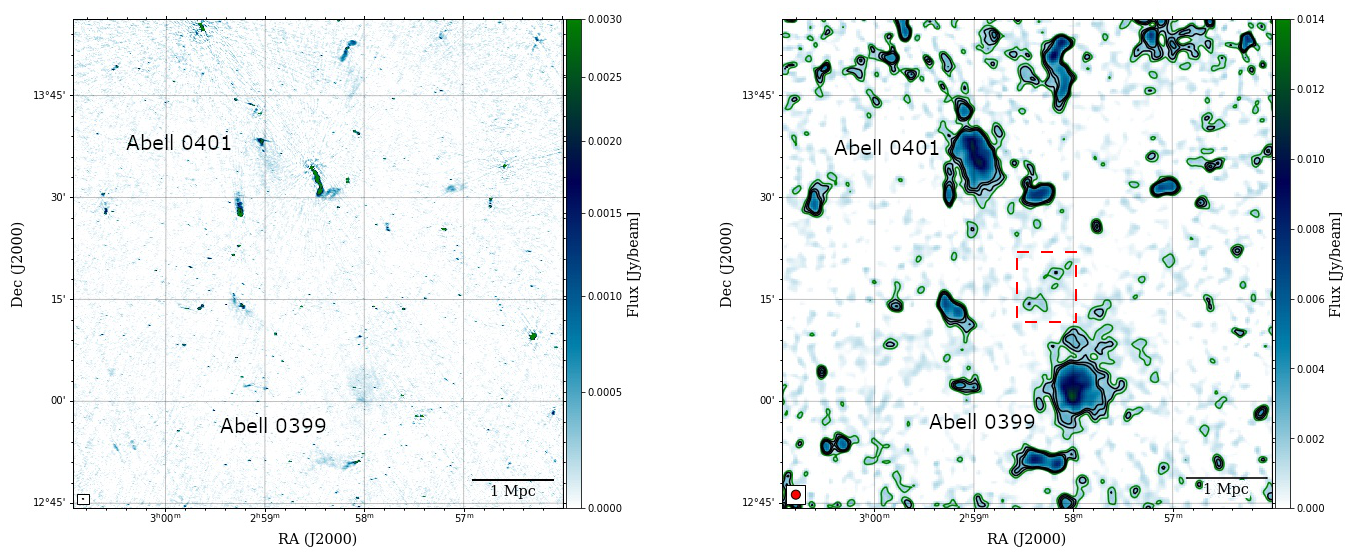}
    \caption{Mosaic radio images at 400~MHz of the A399-A401 clusters pair. \textit{Left panel:} high-resolution ($12^{\prime\prime} \times 5^{\prime\prime}$) mosaic image with $\sigma_{\rm rms} = 50$~$\mu$Jy~beam$^{-1}$ produced with \texttt{Briggs} \texttt{robust = 0} and primary-beam corrected. A hint of diffuse emission from the radio halos is visible, but no emission is detected in the bridge area. \textit{Right panel:} Low-resolution ($80^{\prime\prime}\times80^{\prime\prime}$), compact source-subtracted image with $\sigma_{\rm rms} = 600$~$\mu$Jy~beam$^{-1}$. It is generated by \texttt{Briggs robust=-0.5} and a Gaussian $uv$-taper of $60^{\prime\prime}$. Contour levels start at $2\sigma_{\rm rms}$ (in green) and increase up to $5\sigma_{\rm rms}$ (black). The red-dashed box denotes the region where we find a $2\sigma_{\rm rms}$ detection of a patch of the bridge. The red arrow points to the extension of the A399 radio halo, discussed in \Sec\ref{sec:patch}.}
    \label{fig:high_low_mosaic}
\end{figure*}
\begin{figure*}
    \centering
    \includegraphics[width=1\linewidth]{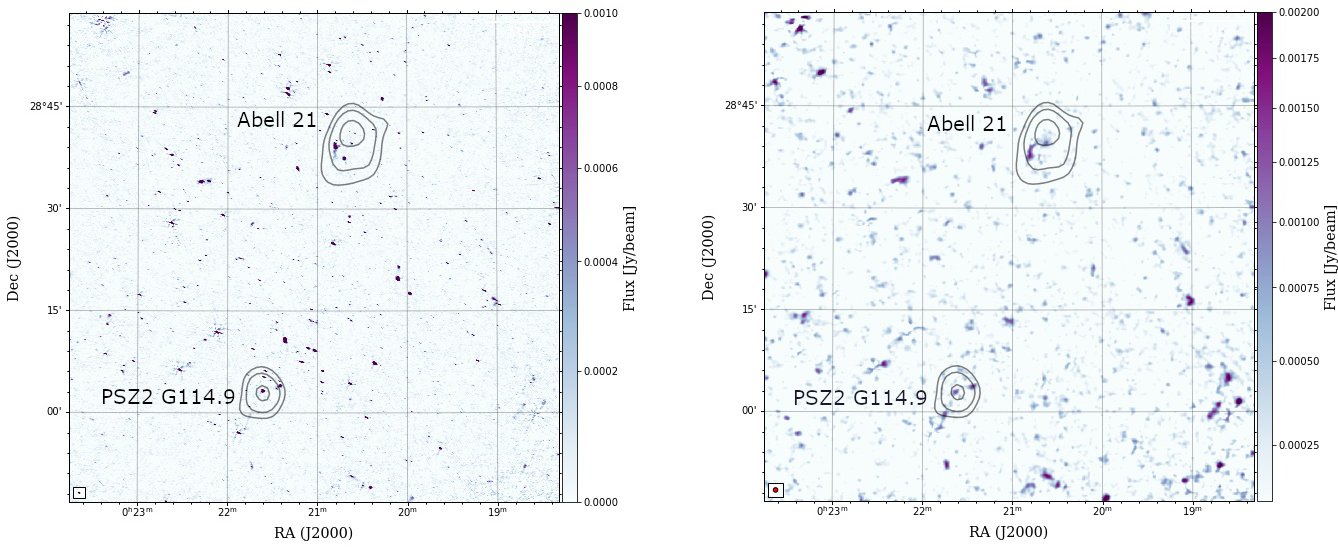}
    \caption{Mosaic radio images at 383~MHz of the A21-G114.9 clusters pair, with overlaid X-ray ROSAT contours. \textit{Left panel:} high-resolution ($15^{\prime\prime} \times 5^{\prime\prime}$) mosaic image with $\sigma_{\rm rms} = 40$~$\mu$Jy~beam$^{-1}$ produced with \texttt{Briggs} \texttt{robust = 0}, showing the compact sources in the field. \textit{Right panel:} Low-resolution ($40^{\prime\prime}\times40^{\prime\prime}$), compact source-subtracted image with $\sigma_{\rm rms} = 230$~$\mu$Jy~beam$^{-1}$. It is generated by \texttt{Briggs robust=0} and a Gaussian $uv$-taper. No diffuse emission is revealed.}
    \label{fig:high_sub_mosaic}
\end{figure*}

\begin{figure}[h!]
    \centering
    \includegraphics[width=1\linewidth]{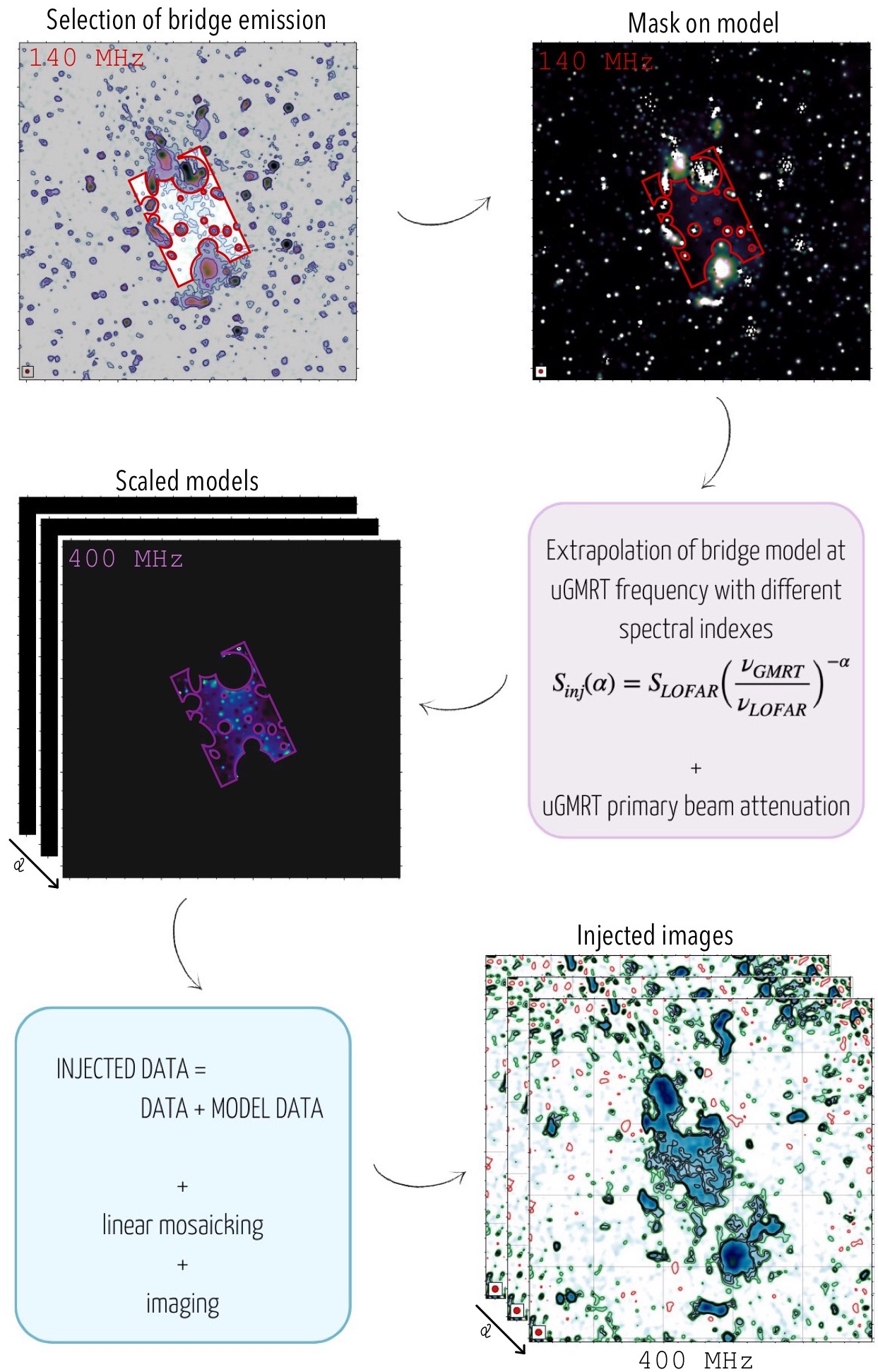}
    \caption{Schematic flowchart of the injection method, described in \Sec\ref{sec:limit}.}
    \label{fig:scheme}
\end{figure}

Before imaging, each pointing of the A399-A401 observation is phase-shifted to a common phase-center ($\text{RA} = 02^{\rm h} \, 58^{\rm m} \, 28^{\rm s}$, $\text{Dec} = +13^{\circ} \, 20^{\prime} \, 18^{\prime\prime}$), and deconvolved individually. We then produced a high resolution ($12^{\prime\prime}\times5^{\prime\prime}$, p.a. $79^{\circ}$) image for each pointing with the same parameters. We adopted a weighting scheme \texttt{Briggs robust=0} \citep{Briggs}, resulting in an rms noise $\sigma_{\rm rms} \sim 70$~$\mu$Jy~beam$^{-1}$, similar in both images, that shows a hint of the radio halos diffuse emission. The choice of this weighting parameters is motivated by the necessity of recovering the diffuse components of the sources in the field. In order to enhance sensitivity on Mpc-scales, we chose to subtract all compact sources in the field, but, as shown in \Fig~\ref{fig:high_sub_mosaic}, we needed to consider also the presence of more extended sources that could contaminate the bridge emission. Thus, we decided to subtract physical scales smaller than 600~kpc: the choice of 600~kpc scale is a tradeoff between the best subtraction of all the compact emission, including the tail of the tailed-radio galaxy in A401 that extends towards the bridge region, while retaining the emission from the radio halos, that extends on scales of approximately 700 kpc. With WSClean, we imaged the field with a \textit{uvmin} of $464\lambda$ to recover only the compact sources, and then subtracted their components from the visibilities.  After producing the high-resolution image with the uv-cut, we carefully inspected the model image to ensure that no components of diffuse emission from the bridge area would be subtracted. For each pointing, we then imaged the source-subtracted data with \texttt{Briggs robust=-0.5} and a Gaussian $uv$-taper to obtain a $80^{\prime\prime}\times80^{\prime\prime}$ resolution image with an rms noise $\sigma_{\rm rms} \sim 800$~$\mu$Jy~beam$^{-1}$. At this point, we can further enhance the diffuse emission present in each pointing, by combining the individual high- and low-resolution images produced with the same parameters into a mosaicked image. This approach is referred to as linear mosaicking \citep[see, e.g.,][]{holdaway_1999}, i.e. the value $\tilde{I}$ of a pixel ${\bf x}$ is the average value among all the $i$ pointings, weighted by the primary beam $P$:
%
\begin{equation}\label{eq:mosaic}
    I(\mathbf{x})=\frac{\sum_{i} P_i(\mathbf{x}) \, I_{i}(\mathbf{x})}{\sum_{i} P_i^{2}(\mathbf{x})},
\end{equation}
here, the summation $i$ is over the pointing centers $x_{i}$, $I_{i}(x)$ is the image produced from the i-th pointing, and $P(x)$ is the uGMRT primary beam pattern\footnote{The uGMRT primary beam shape parameters can be found \hyperlink{http://www.ncra.tifr.res.in/ncra/gmrt/gmrt-users/observing-help/ugmrt-primary-beam-shape}{here}}. 
The results of the linear mosaicking procedure are shown in \Fig~\ref{fig:high_low_mosaic} both for the high- and low-resolution image, with final rms noise of $\sigma_{\rm rms} \sim 50$~$\mu$Jy~beam$^{-1}$ and $\sigma_{\rm rms} \sim 600$~$\mu$Jy~beam$^{-1}$, respectively.

The same general procedure was followed for the A21-G114.9 observation. Each pointing is shifted to a common phase-center ($\text{RA} = 00^{\rm h} \, 21^{\rm m} \, 02^{\rm s}$, $\text{Dec} = +28^{\circ} \, 22' \, 51''$), deconvolved and imaged individually.  For each pointing, we produced a high-resolution ($15''\times5''$, p.a. $68^{\circ}$) image with a weighting scheme \texttt{Briggs robust=0}, and a resulting rms noise $\sigma_{\rm rms} \sim 50$~$\mu$Jy~beam$^{-1}$. In the high-resolution images there is no evidence of diffuse emission, only compact sources are visibile. To investigate the presence of diffuse radio emission corresponding to the SZ detection of the filament reported in \cite{bonjean2018}, we subtracted all compact sources in the field, and then proceed with low-resolution imaging.  
For each pointing, we imaged the source-subtracted data with \texttt{Briggs robust=0}, and a Gaussian $uv$-taper to have a $40'' \times 40''$ resolution image with a rms noise $\sigma_{\rm rms} \sim 300$~$\mu$Jy~beam$^{-1}$. In this case, since the diffuse emission is not revelead, we adopt a moderate weighting scheme to have a quality image showing the features in the field. Finally, we combined the high- and low-resolution individual images of each pointing in two final mosaicked images, following the same approach as A399-A401 pointings. The resulting images with their final rms noise are shown in \Fig\ref{fig:high_sub_mosaic}, where we can notice how there is no visible detection of diffuse emission in the field.

\section{Results for the A399-A401 pair}\label{sec:a399res}
\subsection{Limit to the bridge spectral index}\label{sec:limit}
\begin{figure*}
    \centering
    \includegraphics[width=1\linewidth]{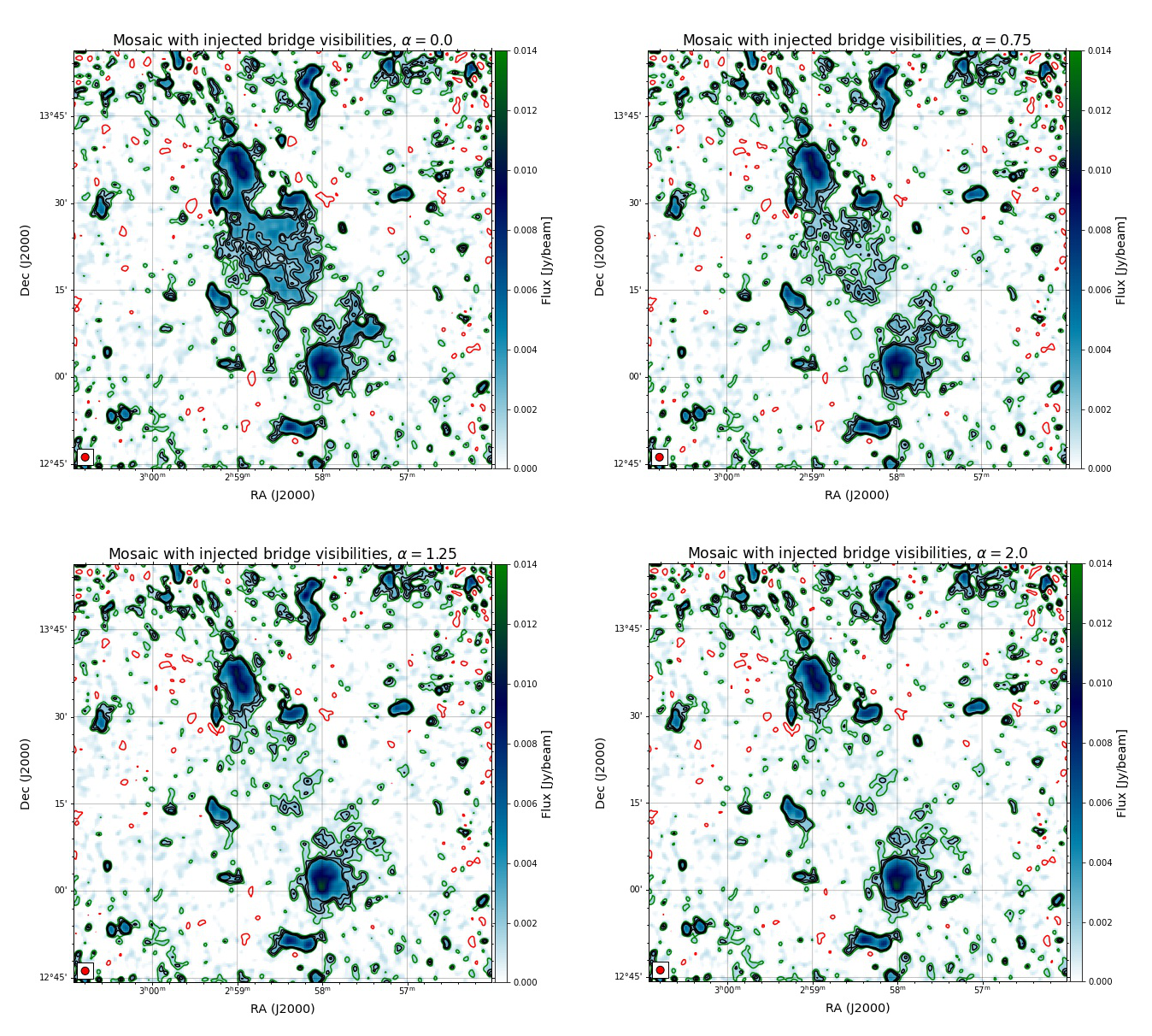}
    \caption{Examples of 400~MHz, uGMRT images where the A399-A401 bridge visibilities are injected, as a function of spectral index. In particular, we show the different contribution of the bridge when its spectral index steepens, from $\alpha=0$ to $\alpha=2.0$. Note how in the first two top panels the bridge is clearly detectable, while approaching the lower limit for the spectral index (left-bottom panel) the emission is less visible, until there is no significant change with respect the image without any injected visibilities. Contour levels are drawn from $2\sigma_{\rm rms}$ (in green) and increase up to $5\sigma_{\rm rms}$ (black). A negative contour level at $-3\sigma_{\rm rms}$ is shown in red.}
    \label{fig:inj_examples}
\end{figure*}
\begin{figure}
    \centering
    \includegraphics[width=1\linewidth]{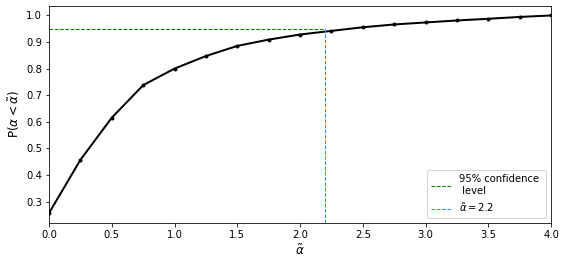}
    \caption{Cumulative distribution function of $R(\alpha)$, normalized to unit area over the interval $0\leq\alpha\leq4$. The horizontal green line marks the $95\%$ probability that the spectral index of the bridge in A399-A401 takes on a value smaller than $\alpha\sim2.2$ (blue vertical line) if the bridge were detected in our uGMRT observations. The non-detection sets a lower limit for the spectral index at $\alpha_{l}>2.2$ with a $95\%$ confidence level. }
    \label{fig:cumulative}
\end{figure}
The presence of a bridge of low-suface brightness radio emission is reported in \cite{govoni2019}, where they detected the diffuse emission between the two galaxy clusters at 140~MHz with LOFAR. 
We are not able to detect the full extension of bridge emission in our 400~MHz uGMRT observations, except for a small patch of emission that we discuss in \Sec~\ref{sec:patch}. Through the non-detection we can, however, place a lower limit on the spectral index of the bridge. The simplest approach would be to use the classical lack-of-detectability criterion, where one places a limit at $3\sigma^{b}_{rms}$, where $\sigma^{b}_{rms}$ is the image rms noise multiplied for the square-root of the number of synthesized beams covering bridge area; in this way, we would find a lower limit on the spectral index $\alpha > 3$. However, this procedure is only appropriate for point-sources, as the noise in interferometric images, generally, does not simply scale with the area, but depends on the baseline disribution, the weighting scheme and the inage fidelity
Therefore, here we will follow a similar procedure to the one first introduced in \cite{venturi2008} for radio halos \citep[see also][]{kale2013, bernardi2016, bonafede2017, duchesne2022}.In particular, we based our method on the work by \cite{nunhokee2021} on the A399-A401 pair at 346~MHz, in order to produce comparable results. They found a lower limit for the spectral index $\alpha > 1.5$ at 96\% confidence level, and given that our observations are more sensitive than the WSRT and we still are unable to detect the bridge emission, we expect to place an even more stringent constraint. 

We refer to this method as \textit{injection}, and a schematic representation of the process is show in \Fig\ref{fig:scheme}. We proceed as follows:
\begin{enumerate}
    \item From the model image of the LOFAR detection at 140~MHz, we created a mask by including only the emission from the bridge. This region is selected to include the emission above the $3\sigma$ contour in the $80^{\prime\prime}$ resolution LOFAR image, inside a  $2\times3$~Mpc box, centered on the bridge ($\text{RA} = 02^{\rm h} 58^{\rm m} 26^{\rm s}$, $\text{Dec} = +13^{\circ} 18^{\prime} 17^{\prime\prime}$, with a position angle of $25^{\circ}$~E of the vertical axis), as defined in \cite{govoni2019}. To make sure we are not including contribution from compact sources in the LOFAR detection, we also masked all sources with emission above the $6\sigma$ contour inside the defined box; 
    \smallskip
    \item As mentioned in \Sec\ref{sec:obsanddata}, our data is divided in six sub-bands of 33.3~MHz each, so we need to extrapolate the bridge model image to each sub-band central frequency $\nu_{n}=[316,348,385,416,449,481]$~MHz with:
    \begin{equation}\label{flux}
        S_{n,\nu_{n}}(x,y,\alpha)=S_{140}(x,y)\left(\frac{\nu_{n}}{\rm 140~MHz}\right)^{-\alpha}
    \end{equation}
    where $S_{n,\nu_{n}}(x,y,\alpha)$ is the flux density of the model image at the frequency $\nu_{n}$ at the pixels $(x,y)$, $S_{140}(x,y)$ is the flux density of the LOFAR model image, and the spectral index varies between $0\leq\alpha\leq4$ with steps of $\Delta\alpha = 0.25$, assuming a uniform spectral index distribution over the source. We discuss the choice of spectral index range in \Sec~\ref{sec:arange}.
    \smallskip
    \item Each extrapolated model image of the bridge is then multiplied with the uGMRT primary beam model to take into account the attenuation of the primary beam in our observation. The final bridge model images are then transformed into visibilities that are injected into our uGMRT source-subtracted, calibrated visibilities of each pointing. We then deconvolved and imaged each pointing separately and then linearly combined them following the same procedure described in \Sec\ref{sec:imaging}. In particular, we produced an $80^{\prime\prime}$ resolution mosaic image at the central frequency of 400~MHz for each spectral index $\alpha$ with the same parameters used to produce the $80^{\prime\prime}$ image from our observations. An example of such images for different spectral indexes is shown in \Fig\ref{fig:inj_examples}. As expected, as the spectral index steepens from $\alpha=0$ to $\alpha=2.0$, the emission of the bridge becomes less and less visible, until is not detectable above the noise level.
    \smallskip
    \item We want to construct a statistical criterion to determine when the bridge is no longer considered detected, i.e. to determine a lower limit to the spectral index of the bridge.  In this sense, the spectral index can be treated as a random variable in the interval $0 \leq \alpha \leq4$, even if we sample it at given values for simplicity. We then define the ratio $R(\alpha)$ as:
    \begin{equation}\label{eq:r}
        R(\alpha)=\frac{\sum_{x=1}^{N}\sum_{y=1}^{N} S_{400} \, (x,y)+S_{n,400} \, (x,y,\alpha)}{\sum_{x=1}^{N}\sum_{y=1}^{N} S_{400}(x,y)} = \frac{S_{400}^{\rm inj}(\alpha)}{S_{400}},
    \end{equation}
    where $S_{400}^{\rm inj}(\alpha)$, defines the flux density of the 400~MHz image with the contribution of the injected visibilities, and $S_{400}$, is the flux density from the 400~MHz image from the uGMRT observations. Both quantities are measured by summing over the $N$ pixels covering the bridge area (see \Fig\ref{fig:scheme}), masking only the area covered by the small patch of emission (red-dashed box in \Fig\ref{fig:high_low_mosaic}, right panel) that we will treat separately in \Sec\ref{sec:patch}. 
    The ratio $R(\alpha)$ is a decreasing function of $\alpha$, it has its maximum value when $\alpha=0$ and approaches unity for increasing spectral index values, i.e. when the bridge spectrum is steeper and therefore, the emission less visible in our injected images. In other words, the ratio $R(\alpha)$ measures how bright, given a certain spectral index value $\alpha$, the injected bridge emission is with respect to the image background.
    While the definition of $R(\alpha)$ is not the formal definition of a distribution function and does not converge for $\alpha\rightarrow-\infty$, it does converge when $\alpha\rightarrow\infty$. We then normalize the integral of $R(\alpha)$ to unity over the spectral index range, so that $R(\alpha)$ effectively represents a probability to detect a spectral index value given our observations.
    \smallskip
    \item We then constructed the cumulative distribution function ($P(\alpha<\tilde{\alpha})$) of $R(\alpha)$ defined as:
    \begin{equation}\label{eq:cpd}
    P(\alpha<\tilde\alpha)=F(\tilde\alpha)=\sum_{0}^{\tilde\alpha} R(\alpha) \, \Delta \alpha.
    \end{equation}
    The cumulative distribution function of $\alpha$ evaluated at $\tilde\alpha$ gives us the probability to observe an emission with a spectral index $\alpha<\tilde\alpha$ in our observations.
    As shown in \Fig\ref{fig:cumulative}, we found that the bridge should be detected in our observations with a probability $P(\alpha<\tilde{\alpha})=95\%$ if $\tilde{\alpha} < 2.2$.
    The non-detection in our observations implies that the spectral index of the bridge has a lower limit of $\alpha_{l}>2.2$ with a 95\% confidence level. Note that the limit value is dependent on the chosen interval range for $\alpha$, as discussed in \Sec\ref{sec:arange}.
\end{enumerate}
This result represents an improvement over the constraints from \cite{nunhokee2021}, due to the higher sensitivity of our observations.
Our lower limit disfavours the shock re-acceleration processes proposed in \cite{govoni2019} as the main mechanism responsible for the bridge emission and is consistent with the predictions from  \cite{brunetti2020}, where the origin of radio bridges is explained with second order Fermi acceleration of electrons interacting with turbulence on over-Mpc scales, resulting in rather steep spectra.

\subsection{The effect of the spectral index range on limit estimates}\label{sec:arange}
 \begin{figure*}[t]
    \centering
    \includegraphics[width=1\linewidth]{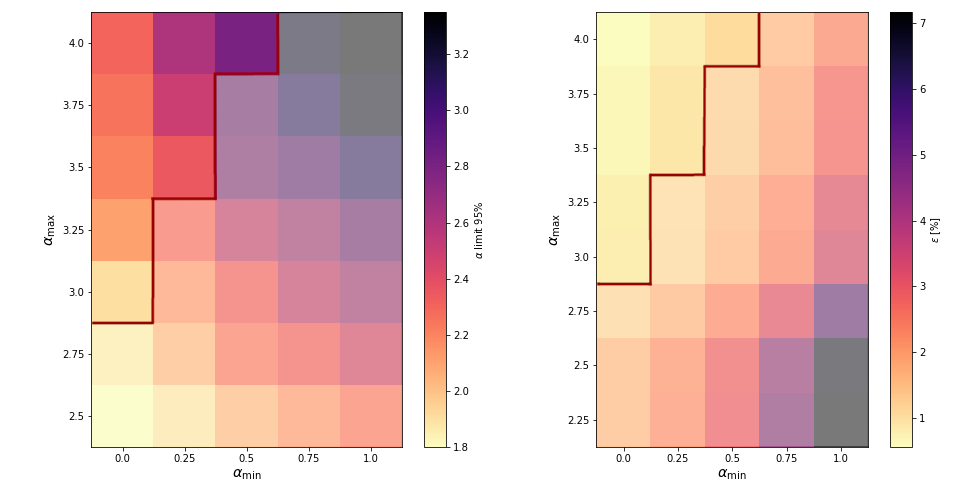}
    \caption{\textit{Left:} Lower limits at 95\% confidence level as a function of spectral index range. The lower and upper bounds of the range can vary between $0 \leq \alpha_{min} \leq 1$ and $0 \leq \alpha_{max} \leq 4$ respectively. Values above the red line satisfy the criterion of convergence of the cumulative distribution (see right panel). We adopted lower limit $\alpha > 2.2$, obtained using the $\alpha_{min} = 0 \leq \alpha \leq4 = \alpha_{max}$ range. \textit{Right:} Convergence of the cumulative probability function with varying spectral index ranges. Values above the red line have $\epsilon < 1\%$, that we use as an acceptance criterion (see \Eq\ref{eq:epsilon}).}
    \label{fig:alpharanges}
\end{figure*}

As presented in \Sec\ref{sec:limit}, we defined the injection procedure to derive the lower limit on the A399-A401 bridge spectral index, by extrapolating the bridge model image from the LOFAR detection to the uGMRT frequency with different spectral index values. We note that the limit value is dependent on the chosen interval range for $\alpha$, thus we tested different ranges to investigate the resulting limit. 
We evaluated the $\alpha_{l}$ value corresponding to the 95\% value of the cumulative distribution (as described in the main text, \Eq\ref{eq:cpd}) as a function of $[\alpha_{min}, \alpha_{max}]$, with $0 < \alpha_{min} < 1$ and $2.25 < \alpha_{max} < 4$, in steps of $\Delta \alpha = 0.25$ (\Fig\ref{fig:alpharanges}, left panel). The $\alpha_{min} > 0$ boundary is motivated by the fact that an inverted spectral index is not expected for synchrotron emission from a bridge-like source. To investigate the convergence of the cumulative distribution, we define $A$ as the area under the curve $R(\alpha)$ (\Eq\ref{eq:r}) calculated between $\alpha_{min}$ and $\alpha_{max}$, before normalization.
We then defined the ratio $\epsilon_i$:
\begin{equation}\label{eq:epsilon}
    \epsilon_i = \frac{(A_{i+1}-A_{i})\times100}{A_{i}},
\end{equation}
 where $i$ runs over the number of $\alpha$ steps.
For each $\alpha_{min}$, $\epsilon_i$ evaluates the percentage difference between the area $A_{i}$ in the interval $\alpha_{min}\leq\alpha\leq\alpha_{max,i}$ and $A_{i+1}$ in the interval $\alpha_{min}\leq\alpha\leq\alpha_{max,i+1}$, where $\alpha_{max,i+1}=\alpha_{max,i}+0.25$. 
In other words, when we fix the value of $\alpha_{min}$, we start with the initial area value $A_{i}$, calculated within the interval defined by $\alpha_{min}$ and $\alpha_{max,i}$. Then, we calculate the area value $A_{i+1}$ within an extended interval that includes $\alpha_{min}$ and an increased upper bound, $\alpha_{max,i}+0.25$. This allows us to compare the percentage difference $\epsilon$, between the area values at each step. We consider the cumulative distribution converging if $\epsilon<1\%$. This is computed for each combination range with $0\leq\alpha_{min}\leq1$, and $2.25\leq\alpha_{max}\leq4$.
 Results are shown in \Fig\ref{fig:alpharanges} (right panel), which shows that $\epsilon$ decreases with increasing $\alpha_{max}$ values. This is somewhat expected as the cumulative distribution function converges for increasing $\alpha_{max}$ values. We notice that $A$ has the strongest dependence upon $\alpha_{min}$, and $\epsilon$ changes up to $\sim 6\%$ across the $0 < \alpha_{min} < 1$ - a small variation anyway. We assumed that estimates of the spectral index lower limit began to converge if $\epsilon < 1\%$, i.e., if the relative variation between the area under the curve $R(\alpha)$ is smaller than 1\% (values above the red line in \Fig\ref{fig:alpharanges}, left panel). We choose to report the case where the convergence is strongest, with $\epsilon \sim 0.5\%$, leading to $\alpha > 2.2$ for $0 \leq \alpha \leq 4$. 

\subsection{Detection of a patch of bridge emission}\label{sec:patch}

As mentioned in \Sec\ref{sec:imaging}, we observe a $2\sigma_{\rm rms}$ level patch of emission in the bridge area in the $80^{\prime\prime}$ resolution image, encompassed by the red-dashed box shown in \Fig\ref{fig:high_low_mosaic}. There were no compact sources in the location corresponding to this region prior to the source subtraction process. Under the assumption that this patch represents a part of the bridge, and that the spectral index is likely variable across the bridge area, this region could present a spectral index flat enough to be detectable in our observations.

Within the $2\sigma_{\rm rms}$ level contours of the uGMRT image, we measure flux densities of $S_{\rm 400~MHz} = 8.7 \pm 1.7$ mJy and $S_{\rm 140~MHz} = 26.7 \pm 3.7$ mJy, leading to a spectral index value for the patch of $\alpha_{p}= 1.07 \pm 0.23$. \\
The uncertainty on the flux density measurements is estimated as: 
\begin{equation}
    \sigma_{S}= \sqrt{(S\times f)^{2}+N_{b}\times (\sigma_{\rm rms})^{2}},
\end{equation}
where $f=0.1$ is the absolute flux-scale uncertainty \citep{chandra2004}, $N_{b}$ the number of beams covering the source, $\sigma_{\rm rms}$ the rms noise sensitivity of the map and $S$ is the measured flux density of the source.\\
As expected, the spectral index of this emission is significantly flatter than our lower limit, thus could be revealing a small part of the bridge in our observations. \\
The interpretation of this $2\sigma_{\rm rms}$ level patch of emission is not definitive at this time, but it is possible to make some considerations based on the models and detections available in the literature. A physical scenario that could explain the presence of flatter emission patches is one of the predictions from the turbulent re-acceleration model proposed in \cite{brunetti2020}. In fact, with their simulations they show how the volume filling factor of the bridge emission should be larger at LOFAR frequencies, resulting in a smoother emission, but at higher frequencies the emission is predicted to be dominated by a clumpy contribution from smaller, more turbulent regions. Moreover, they show that even during the early stages of a merger between two systems, the dynamics of the collapse can drive weak shocks into the inter-cluster medium, resulting in an additional compression of the population of turbulent re-accelerated electrons, increasing the radio brightness in these location.

\section{Results for the A21-G114.9 pair}\label{sec:results_a21psz}
\begin{figure*}
    \centering
    \includegraphics[width=1\linewidth]{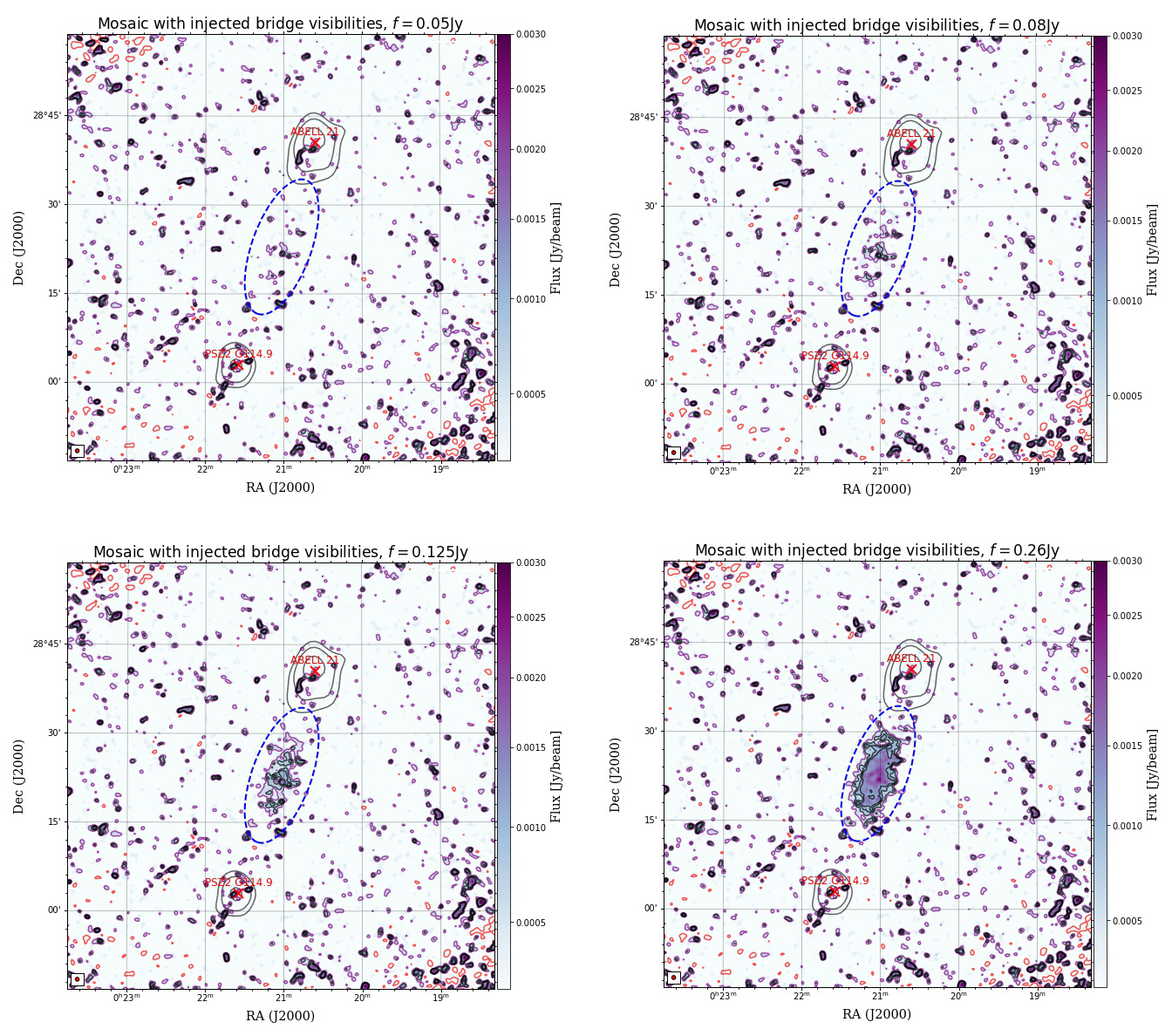}
    \caption{Examples of 383~MHz, uGMRT images of A21-G114.9 where the bridge visibilities are injected, as a function integrated flux density $f$. X-ray ROSAT contours are shown in yellow, and the location of the X-ray peak is marked with yellow crosses. In particular, we show the different contribution of the injected bridge emission when increasing its flux density, from $f=50$mJy to $f=260$mJy. The visibilities are injected inside a 2D model Gaussian (blue dashed ellipse) with semi-major axis of $12'$ ($\sim 2.5$Mpc) and semi-minor axis of $5'$ ($\sim 500$kpc). Note how in the first two top panels the bridge emission is not significantly detected, while approaching the two values found for the upper limit on the flux density (bottom panels) the emission is continuous and detected over $2\sigma_{rms}$. Contour levels are drawn from $2\sigma_{\rm rms}$ (in purple) to $5\sigma_{\rm rms}$ (in black). A negative contour level at $-3\sigma_{\rm rms}$ is shown in red.} 
    \label{fig:inj_flux}
\end{figure*}
\begin{figure*}[t]
    \centering
    \includegraphics[width=1\linewidth]{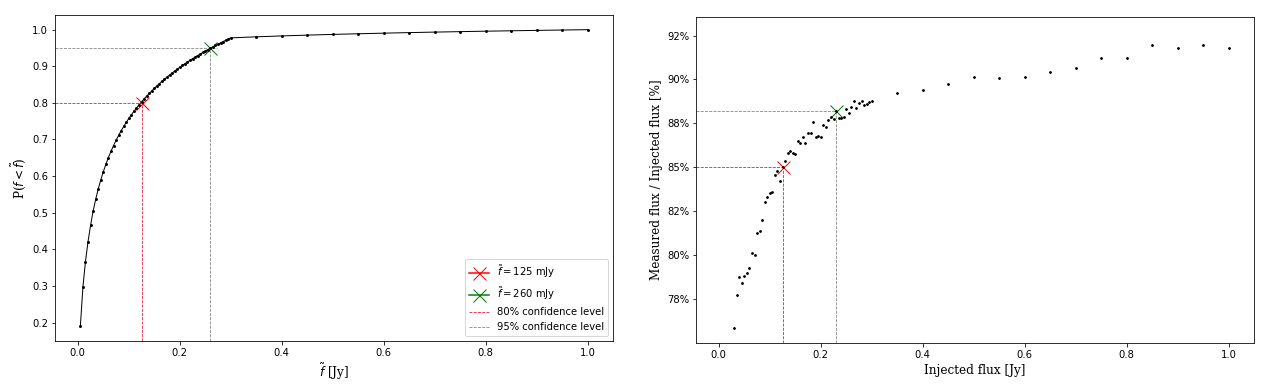}
    \caption{\textit{Left panel:} Cumulative distribution function of $R(f)$, normalized to unit area over the interval $5\text{mJy}\leq f\leq1$Jy. The horizontal grey line marks the $95\%$ probability that the bridge in A21-G114.9 has a flux density lower than $\tilde{f}=260$mJy (green cross), since it is not detected in our observations. This sets an upper limit on its flux density at $f_{u}^{1}< 260$ with a $95\%$ confidence level. With a second criterion based on the extension and continuity of the bridge emission (see \Sec\ref{sec:results_a21psz}), we find a lower value for the upper limit at $f_{u}^{2}< 125$mJy. From the cumulative function, there is a 80\% probability that the bridge emission is lower than 125~mJy (red cross). \textit{Right panel:} plot of the ratio (in percentage) between the measured flux density and the injected flux density, with varying injected flux density. The green cross corresponds to the recovered percentage of the 260mJy injected bridge emission ($\sim 88\%$), and the red cross corresponds to the recovered percentage of the 125mJy injected bridge emission ($85\%$). We notice that the fractional recovered flux increases at increasing injected flux density, converging around $\sim92\%$. The injected flux density lost in this procedure is never higher than $\sim 23\%$. }
    \label{fig:cumulative_ratio}
\end{figure*}

In this case, we need to assess how deep our observation should be to detect possible diffuse emission from the filament. This is the first time the A21-G114.9 pair has been observed at radio frequencies, therefore as opposed to A399-A401, we can we can only place a limit on the bridge flux density assuming a model for the bridge emission. We define the morphology and the profile of the mock radio bridge to fit the observations of the most in-depth bridge study \cite{govoni2019}, and to follow the elongated shape of filamentary emission we would expect from the SZ detection. We detail below the steps we have followed:
\begin{enumerate}
    \item For the mock bridge brightness profile, we assume a two-dimensional elliptical Gaussian profile:
    \begin{equation}
    \begin{split}
        I(x,y)=A\exp[&(-(a(x-x_{0})^{2}+ 
        \\&2b(x-x_{0})(y-y_{0})+c(y-y_{0})^2)],
    \end{split}
    \end{equation}
    where
    \begin{equation}
    \begin{split}
        a & =\frac{cos^{2}\theta}{2\sigma^{2}_{x}}+\frac{sin^{2}\theta}{2\sigma^{2}_{y}}, \\
        b & =-\frac{sin2\theta}{4\sigma^{2}_{x}}+\frac{sin2\theta}{4\sigma^{2}_{y}},\\
        c & =\frac{sin^{2}\theta}{2\sigma^{2}_{x}}+\frac{cos^{2}\theta}{2\sigma^{2}_{y}},
    \end{split}
    \end{equation}
    and where $\sigma$ is the Gaussian standard deviation and $\theta$ is the rotation angle.
    The  2D Gaussian model is centered in $(x_0,y_0)$=(RA,Dec)=($00^{\rm h} \, 21^{\rm m} \, 02^{\rm s}, +28^{\circ} \, 22^{\prime} \, 51^{\prime\prime}$). The semi-major axis is $\sigma_{y}=12^{\prime}$, the semi-minor axis is $\sigma_{x}=5^{\prime}$, and the ellipse is rotated of $\theta=20^{\circ}$ W of the vertical axis. We scale the amplitude $A$ so that the integrated flux density of the injected mock bridge varies between $5~\text{mJy}-300~\text{mJy}$ with increasing steps of $5~\text{mJy}$, and between $300~\text{mJy}-1~\text{Jy}$ with increasing steps of $50~\text{mJy}$.
    \smallskip
    \item Each Gaussian model is then multiplied with the uGMRT primary beam model to take into account the attenuation of the primary beam in our observations. 
    \smallskip
    \item The final mock bridge models are transformed into visibilities and injected into our uGMRT source-subtracted calibrated data of each pointing. Then we follow the same procedure described in \Sec\ref{sec:imaging} to produce $40^{\prime\prime}$ resolution mosaic images at the central frequency of 383~MHz for each model flux density $f$. Examples of the resulting injected images with different injected flux densities are shown in \Fig\ref{fig:inj_flux}.
    \smallskip
    \item We define the ratio $R(f)$ as:
    \begin{equation}
    R(f)=\frac{\sum_{x=1}^{N}\sum_{y=1}^{N} S_{383}(x,y)}{\sum_{x=1}^{N}\sum_{y=1}^{N} S_{383} \, (x,y)+S_{n,383} \, (x,y,f)} = \frac{S_{383}}{S_{383}^{\rm inj}(f)}
    \end{equation}
    where $S_{383}$ defines the flux density of the 383~MHz image with no injected visibilities, and $S_{383}^{\rm inj}(f)$ is the flux density from the 383~MHz image with the contribution of the injected visibilities. Both quantities are measured by summing over the N pixels covering the bridge area within the Gaussian ellipse. The ratio $R(f)$ is a decreasing function of the injected flux density $f$, hence in the limit $f\rightarrow\infty$ (increasing injected flux), $R(f)\rightarrow 0$. This implies that there is a value of the injected flux $f$, for which the bridge emission should be significantly detected in our 383~MHz observation (i.e., $S_{383}^{\rm inj}(\geq f_{u})> S_{383}$). Since the bridge is not detected, this sets an upper limit on its flux density at $f_{u}$. 
    \smallskip
    \item Following the same procedure of the A399-A401 case, we determine the flux density upper limit by constructing the cumulative distribution function ($P(f<\tilde{f})$, as defined in \Eq\ref{eq:cpd}) of $R(f)$, normalized to unit area over the defined interval for $f$. As shown in \Fig\ref{fig:cumulative_ratio} (left panel), we found that $P(f<\tilde{f})=95\%$ for $\tilde{f}\sim260$mJy. This means that there is a 95\% probability that the bridge emission is lower than 260~mJy, otherwise it would be clearly detected in our uGMRT observations. This sets an upper limit on the bridge flux density at $f_{u}^{1}< 260$mJy with a 95\% ($\sim 2\sigma$) confidence level. However, a visual inspection of the image with 260~mJy injected flux (see \Fig\ref{fig:inj_flux}), shows that the bridge would be detected at $5\sigma_{\rm rms}$. As we are assessing for the first time a procedure to place upper limits to the bridge  emission, we have evaluated a second criterion for detection, based on the extension and continuity of the injected diffuse emission, as already done for radio halos (see, e.g., \citealt{bonafede2017}).
    \smallskip
    \item We measure $L_{f}^{2\sigma}$, the largest detectable size of continuous injected emission above $2\sigma_{\rm rms}$, for each image of injected flux density $f$. We consider the bridge detected when the emission above $2\sigma_{\rm rms}$ is continuous for at least the extent of the semi-major axis of the model Gaussian ellipse ($L_{f}^{2\sigma}\geq \sigma_{y}$). With this criterion, we find that we would consider the bridge detected when $f\geq125$mJy. If we define $A_{\rm tot}$ as the total area of the model Gaussian ellipse over which we performed the injection, the area covered by the emission above $2\sigma_{\rm rms}$ in the $f=125$mJy image corresponds to the $28\%$ of $A_{\rm tot}$. 
    This second method sets a lower value for the upper limit, $f_{u}^{2}<125$mJy. This result is in agreement with the visual inspection of the images, and would be equivalent to the 80\% confidence level from the cumulative probability function (see \Fig\ref{fig:cumulative_ratio}, left panel). 
 \end{enumerate}   
This procedure represents the first attempt of adapting to radio bridges the pre-existing injection method introduced in \cite{venturi2008} for upper limits on radio halos. \\
These results are dependent on the model we adopted to describe the possible bridge emission.
Moreover, given the very few detection of radio bridges so far, the modelling of the morphology and surface brightness of mock emission on such large scales is subjected to some arbitrary choices. In comparison with the previous methods, this process presents an improvement by associating a confidence level to the upper limit value, and we are able to compare the results from a second criterion based on the continuity of the recovered emission.

\subsection{Fractional recovered flux density}

As already noticed when the injection procedure was first introduced for radio halos \citep{venturi2008}, we expect that the measured flux density of the mock bridge can be different than the injected flux density, as the faintest components may not be found during the imaging process. \\
As a final consideration, to report the measured value of the flux density upper limit, we evaluated how much of the injected model flux density is effectively recovered in the images we produced. In \Fig\ref{fig:cumulative_ratio} (right panel) we have plotted the fractional recovered flux density with the injected flux density (i.e. the ratio measured flux over injected flux). We can see that the percentage of flux lost in the injection procedure is always smaller than the $23\%$ of the injected flux. In particular, for the upper limits found in the two different method explained above, we found that the measured value of flux is 
    \begin{equation}
    \begin{split}
        f^{\rm meas,1}&=f^{1}*0.88=229 \text{mJy} \\  
        f^{\rm meas,2}&=f^{2}*0.85=106 \text{mJy},
    \end{split}
    \end{equation}
    as shown in \Fig\ref{fig:cumulative_ratio} (right panel)  with the green and red cross, respectively.
    Hence, the resulting upper limits on the measured flux density are:
    \begin{equation}
    \begin{split}
        f^{\rm meas,1}_{u}& <229 \text{mJy}, \\  
        f^{\rm meas,2}_{u}& <106 \text{mJy}.
    \end{split}
    \end{equation}

As expected, the loss effect is generally more important at lower flux densities, where the faintest components of the mock bridge on larger scales can result below the noise level. At increasing flux density, the effect is less severe, and the fractional recovered flux converges around $\sim92\%$.
%

\section{Conclusions}

In this work, we have analysed uGMRT data of two unique systems of early-stage merging galaxy clusters, where their connection along a filament of the cosmic-web is supported by a significant SZ effect detection by the Planck satellite \citep{bonjean2018, planck2013, planck2016}. The A399-A401 pair was already studied with LOFAR, detecting extended diffuse emission in the region inter-cluster \citep{govoni2019}; the A21-G114.9 pair was unexplored at radio frequencies. Our results can be summarized as follows:
\begin{enumerate}
    \item For the A399-A401 system we are not able to detect the full extension of the bridge emission in our 400~MHz observations. We follow the injection method (\citealt{venturi2008, nunhokee2021}) to inject the model visibilities of the detected bridge emission at 140~MHz \citep{govoni2019} in our observations, scaling the flux density with different values of spectral indices. We find that the bridge would be detected in our observations if its spectral index were flatter than 2.2 with a 95\% confidence level, setting a lower limit at $\alpha_{l}>2.2$. This result allows us to test the theoretical models for the bridge origin, disfavouring the shock scenario proposed in \cite{govoni2019}, and is instead consistent with the global predictions of the turbulent (re-)acceleration model of electrons of \cite{brunetti2020}.
    \smallskip
    \item We observed a $2\sigma_{\rm rms}$ significance patch of emission in the bridge area. Under the assumption that this could represent a part of the bridge emission, for this patch we find a spectral index value of $\alpha_{p}= 1.07 \pm 0.23$, significantly flatter than our limit. This result can indicate a variable spectral index distribution across the bridge area. 
    \smallskip
    \item For the A21-G114.9 system, we do not recover any diffuse emission in our 383~MHz observations. We follow a similar injection procedure, but in this case we place an upper limit on the flux density of the bridge emission by assuming an elliptical Gaussian model for the description of the mock bridge surface brightness profile. 
    From the injection, we find a flux density upper limit at  $f_{u}^{1}<260$mJy with 95\% confidence level.
    \smallskip
    \item We propose a second criterion for placing the upper limit, based on the morphology and continuity of the injected emission recovered in the images. In particular, we consider the bridge detected when the emission is above $2\sigma_{\rm rms}$ is continuous for at least the extent of the semi-major axis of the model Gaussian ellipse that defines the injected mock bridge. With this criterion, we find that the upper limit can be placed at $f_{u}^{2}<125$mJy, in agreement with the visual inspection of the images and equivalent  to a 80\% confidence level from the cumulative probability function. 
    \smallskip
    \item We have investigated how much of the injected flux is effectively recovered at the end of the injection procedure. We find that the percentage of recovered flux increases with the injected flux and converges around 92\%, and with our methodology, we consider it unlikely that more than 23\% of the injected flux is lost.

\end{enumerate}

The limits that we have derived represent an important constrain for the spectral characterisation of the emission in radio bridges. The large-scale extension, low-surface brightness and steep spectra that we expect from the theoretical models and from the few present observations, pose a challenge for multi-frequency detection. However, we have defined in this work a procedure to derive upper limits on their flux density that can be applied to more systems in future observations, that will lead to a more comprehensive view of the radio bridges properties and a statistical assessment of their occurrence. 

\begin{acknowledgements}
The authors would like to acknowledge the help and contribution of F. Ubertosi in producing the X-ray images.
AB, CJR acknowledge financial support from the ERC Starting Grant `DRANOEL', number 714245.
\end{acknowledgements}

%
%

\bibliographystyle{aa}
\bibliography{bibliography}

\end{document}